\documentclass[aps,preprint,preprintnumbers,amsmath,amssymb,amsfonts,groupedaddress,superscriptaddress,prl,showpacs,showkeys,floatfix]{revtex4-1}

\usepackage{graphicx}
\usepackage{subfigure}
\usepackage{bm}% bold math
\usepackage{color}
\usepackage{array}
\usepackage{epstopdf}

\newcommand{\pr}{\prime}
\newcommand{\Ga}{\Gamma}

\begin{document}

\title{Visualization of electronic topology in ZrSiSe by scanning tunneling microscopy}

\author{Kunliang Bu}
\affiliation{Department of Physics, Zhejiang University, Hangzhou 310027, China}
\author{Ying Fei}
\affiliation{Department of Physics, Zhejiang University, Hangzhou 310027, China}
\author{Wenhao Zhang}
\affiliation{Department of Physics, Zhejiang University, Hangzhou 310027, China}
\author{Yuan Zheng}
\affiliation{Department of Physics, Zhejiang University, Hangzhou 310027, China}
\author{Jianlan Wu}
\affiliation{Department of Physics, Zhejiang University, Hangzhou 310027, China}
\author{Fangchu Chen}
\affiliation{Key Laboratory of Materials Physics, Institute of Solid State Physics, Chinese Academy of Sciences, Hefei 230031, China}
\affiliation{University of Science and Technology of China, Hefei 230026, China}
\author{Xuan Luo}
\affiliation{Key Laboratory of Materials Physics, Institute of Solid State Physics, Chinese Academy of Sciences, Hefei 230031, China}
\author{Yuping Sun}
\affiliation{Key Laboratory of Materials Physics, Institute of Solid State Physics, Chinese Academy of Sciences, Hefei 230031, China}
\affiliation{High Magnetic Field Laboratory, Chinese Academy of Sciences, Hefei 230031, China}
\affiliation{Collaborative Innovation Center of Advanced Microstructures, Nanjing 210093, China}
\author{Qiunan Xu}
\affiliation{Beijing National Laboratory for Condensed Matter Physics and Institute of Physics, Chinese Academy of Sciences, Beijing 100190, China}
\author{Xi Dai}
\affiliation{Beijing National Laboratory for Condensed Matter Physics and Institute of Physics, Chinese Academy of Sciences, Beijing 100190, China}
\affiliation{Department of Physics, Hong Kong University of Science and Technology, Clear Water Bay Road, Kowloon, Hong Kong}
\author{Yi Yin}
\email{yiyin@zju.edu.cn}
\affiliation{Department of Physics, Zhejiang University, Hangzhou 310027, China}
\affiliation{Collaborative Innovation Center of Advanced Microstructures, Nanjing 210093, China}

\begin{abstract}
As emerging topological nodal-line semimetals, the family of ZrSiX (X = O, S, Se, Te) has attracted broad interests in condensed matter physics due to
their future applications in spintonics.
Here, we apply a scanning tunneling microscopy (STM) to study the structural symmetry and electronic topology of ZrSiSe.
The glide mirror symmetry is verified by quantifying the lattice structure of the ZrSe bilayer
based on bias selective topographies. The quasiparticle interference analysis is used to identify
the band structure of ZrSiSe. The nodal line is experimentally determined at $\sim$ 250 meV above the Fermi level.
An extra surface state Dirac point at $\sim$ 400 meV below the Fermi level is also determined.
Our STM measurement provides a direct experimental evidence of the nodal-line state in the family of ZrSiX.

\end{abstract}

\maketitle

The topology of electronic bands is closely correlated with intrinsic
symmetries in topological materials~\cite{chiu2016classification}. The three dimensional (3D) Dirac semimetals host a fourfold degenerate Dirac point,
which is protected by spatial inversion symmetry, time reversal symmetry and additional threefold  or fourfold rotational symmetry along the $z$-axis~\cite{liu2014discovery, wang2012dirac}.
If one symmetry is broken, the spin-doublet degeneracy of the bands is removed and the Dirac point
is changed to the twofold degenerate Weyl point, leading to a Weyl semimetal~\cite{weng2015weyl, lv2015experimental, xu2015discovery, inoue2016quasiparticle}.
In contrast, the topological nodal-line semimetals host a loop of Dirac points in the momentum space,
which has recently been  predicted theoretically and verified experimentally~\cite{xu2015two, bian2016topological, hu2016evidence}.
The formation of a nodal line requires extra symmetries, such as
mirror reflection symmetry~\cite{bian2016topological} or glide mirror symmetry~\cite{xu2015two}.
The appearance of Dirac or Weyl points near the Ferimi level gives rise to exotic electronic properties,
such as large magnetoresistance~\cite{ali2016butterfly, wang2016evidence, lv2016extremely, singha2017large},
high carrier density~\cite{hu2017nearly} and mobility~\cite{wang2016evidence, ali2016butterfly, sankar2017crystal}.
The nodal-line semimetals are thus good candidates of spintronics for both fundamental research and future applications.

In a previous study, a nodal line was observed in the band structure of PaTaSe$_2$~\cite{bian2016topological}.
Due to interference of other bands, the investigation of nodal-line Dirac fermions is difficult around the Fermi level.
In a different family of ZrSiX (X = O, S, Se, Te) semimetals with glide mirror symmetry,
the nodal line is theoretically predicted~\cite{xu2015two}. The calculation shows that the Dirac cone is
linearly dispersed in a large energy range ($\sim$ 2 eV), without interference of other bands.
Through the measurement of the band structure below the Fermi level, angle-resolved photoemission spectroscopy (ARPES)
has probed the linear band dispersions of ZrSiS and ZrSiSe~\cite{schoop2016dirac, neupane2016observation, hosen2017tunability}.
However, the theoretical prediction of the nodal line is above the Fermi level so that ARPES cannot make a direct
measurement. Instead, scanning tunneling microscope (STM) is a powerful tool to detect both the topography
and local density of states (LDOS), which provides a transparent view of microscopic properties.
A quasiparticle interference (QPI) technique can be used to extract the band structure
in a broad energy above and below the Fermi level~\cite{hoffman2002imaging}.
The previous STM measurements on ZrSiS however did not really determine the nodal-line state due
to their limitations in data acquisition and analysis~\cite{lodge2017observation, butler2017quasiparticle}.

In this paper, we take the STM measurement on ZrSiSe.
Our study resolves a bias selective topography and precisely identifies an atomic shift between
Zr and Se sublattices, giving an evidence of the glide mirror symmetry in ZrSiSe. The QPI analysis visualizes
the linear band dispersion, which determines a nodal line located at $\sim250$ meV above the Fermi level
and a Dirac point located at $\sim400$ meV below the Fermi level.
Our measurement is thus the first STM determination of the nodal-line state in the family of ZrSiX.

High-quality single crystals of ZrSiSe in our experiment are grown by the chemical vapor transport method.
STM measurements are carried out in a commercial ultra-high vacuum system~\cite{zheng2017study}.
The samples are cleaved {\it in situ} at liquid nitrogen temperature and immediately inserted into the STM head.
An electrochemically etched tungsten tip is treated with the field emission on a single crystalline of Au(111) surface.
All data are acquired at liquid helium temperature ($\sim$~4.5 K).

\begin{figure}%[tp]
\centering
\includegraphics[width=0.5\columnwidth]{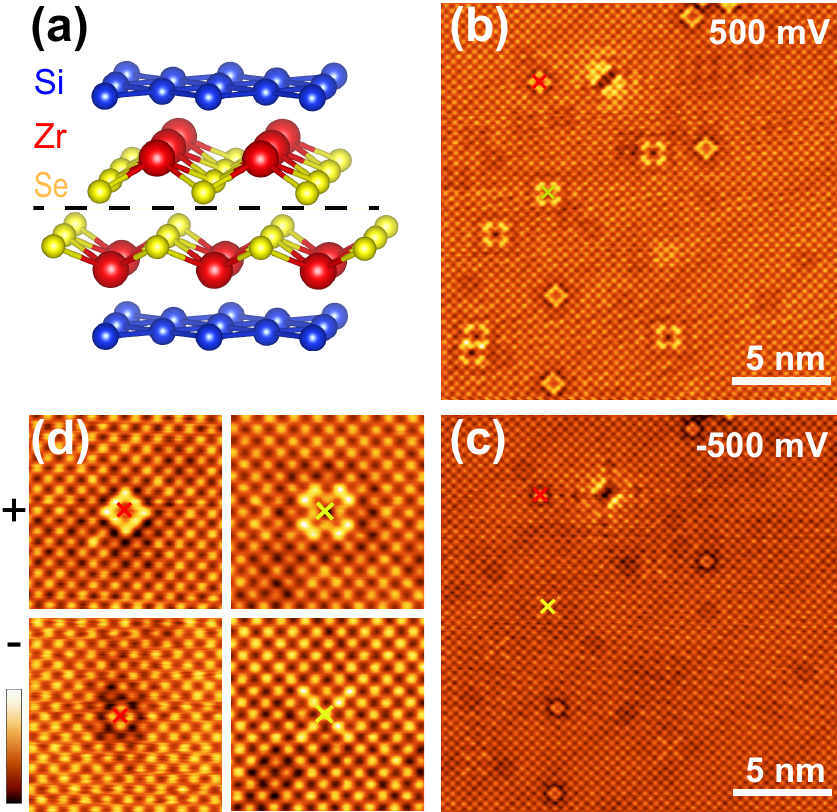}
\caption{(a) A schematic diagram of ZrSiSe. Two 20 nm $\times$ 20 nm topographic images
of the same FOV under the tunneling current of $I=400$ pA and the bias voltages of (b) $V_{\mathrm b}=500$ mV and (c) $V_{\mathrm b}=-500$ mV.
The local topographic images of two defects labelled by the red and yellow crosses are enlarged in (d), where
the upper and lower panels correspond to the positive and negative bias voltages, respectively.}
\label{fig_n01}
\end{figure}

The crystal structure of ZrSiSe is in the space group of $P$4/$\it{nmm}$~\cite{xu2015two}.
As shown in Fig.~\ref{fig_n01}(a), each Si square layer is sandwiched by two sets of ZrSe bilayers.
The crystal is cleaved in between two adjacent ZrSe bilayers and a Se square layer is exposed to be the surface plane.
Figures~\ref{fig_n01}(b) and~\ref{fig_n01}(c) display topographies under two opposite bias voltages in the same field of view (FOV).
The detected lattice is shifted from the top to hollow sites when the bias voltage is switched.
As an illustration, we present enlarged images of two defects in Fig.~\ref{fig_n01}(d).
Under the positive bias voltage [Fig.~\ref{fig_n01}(d), upper panels], the centers of the diamond and cross shaped defects are at the top
and hollow sites, respectively.
Under the negative bias voltage [Fig.~\ref{fig_n01}(d), lower panels], these two centers are switched to their opposite sites.
Two different sublattices are detected in the STM, each selected by a specific bias voltage polarity.

\begin{figure}[tp]
\includegraphics[width=0.5\columnwidth]{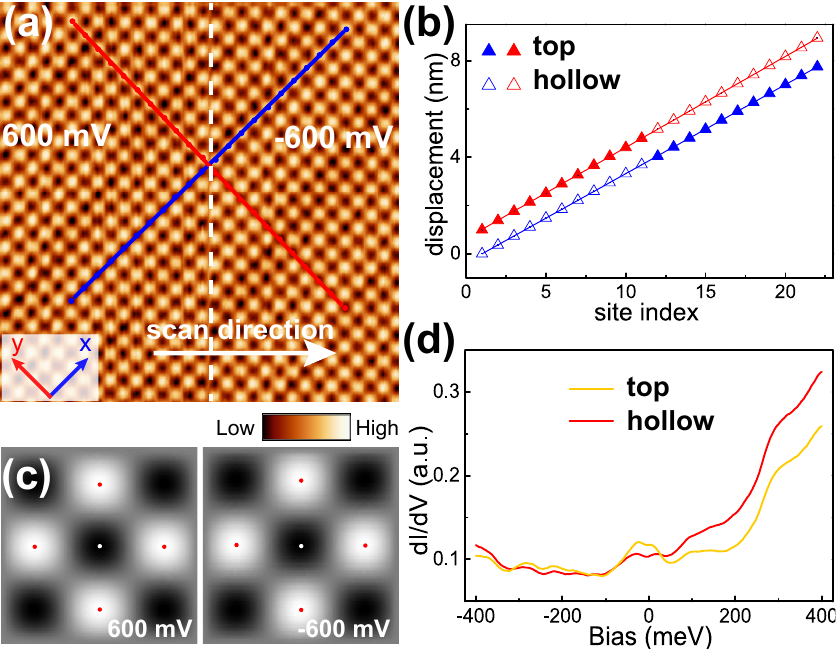}
\caption{(a) A 8 nm $\times$ 8 nm topographic image %($I=500$ pA)
where the bias voltage is switched from
$V_{\rm b}=600$ mV to $V_{\rm b}=-600$ mV when crossing the dashed line.
(b) The relative displacement of each site (top or hollow) along the blue and red linecuts in (a).
Data are offset vertically for clarity. The up-triangles represent the experimental result and the solid lines are from a linear fitting.
The solid and open up-triangles refer to the top and hollow sites, respectively.
(c) The two `supercell'  images obtained for the left and right topographic images in (a).
The bright and dark spots refer to the averaged top and hollow sites, with
red and white dots labeling their centers.
(d) The average $dI/dV$ spectra at the top (orange) and hollow (red) sites under the negative bias voltage. }
\label{fig_n02}
\end{figure}

To further explore the atomic structure of ZrSiSe, we perform a designed experiment
on a clean surface. As shown in Fig.~\ref{fig_n02}(a),  the topography is scanned along the left-to-right direction.
The applied bias voltage is initially positive and suddenly switched to a
negative value at an intermediate position [Fig.~\ref{fig_n02}(a), dashed line].
With respect to this switching line, the left and right topographies are shifted due to the change
of the bias voltage.
Two linecuts (blue and red) along the $x$- and $y$-directions are selected as a demonstration.
The red line crosses the top sites under the positive bias and then the hollow sites under the negative bias.
In Fig.~\ref{fig_n02}(b), we record the sequence of these sites and plot their relative displacements,
which are in a perfect linear relation with the site index.
The same behavior is observed for the sites along the blue line. The hollow sites
under the negative bias  are thus extended from the top sites under the positive bias, and vice versa.

For each of the left and right topographies,
a `supercell' technique~\cite{lawler2010intra, fujita2014simultaneous, zeljkovic2015dirac}
is applied to extract an averaged image with a significantly reduced error.
The ($\pm a_0/2$, $\pm a_0/2$) spatial displacement
with the lattice constant $a_0=$ 3.62~\AA~is precisely determined between neighboring
top and hollow sites for both topographies [Fig.~\ref{fig_n02}(c)].
Under a given bias voltage, the top and hollow
sites form two different sublattices, attributed to two planes of the ZrSe bilayer.
To identify their components,
we average the $dI/dV$ spectra over the top and hollow sites separately.
The local density of unoccupied states at the hollow sites is consistently larger than that
at the top sites [Fig.~\ref{fig_n02}(d)].
The 5$d$ orbitals of Zr atoms are highly unoccupied while the 4$p$ orbitals of Se atoms are highly filled.
Thus, the sublattices of the top and hollow sites with the negative bias correspond
to the Se and Zr layers, respectively. The opposite result can be obtained for the positive bias.
Our topography measurement visualizes the atomically resolved structure of the ZrSe bilayer,
which obeys a key requirement of the glide mirror symmetry in ZrSiSe.

\begin{figure}[tp]
\includegraphics[width=0.75\columnwidth]{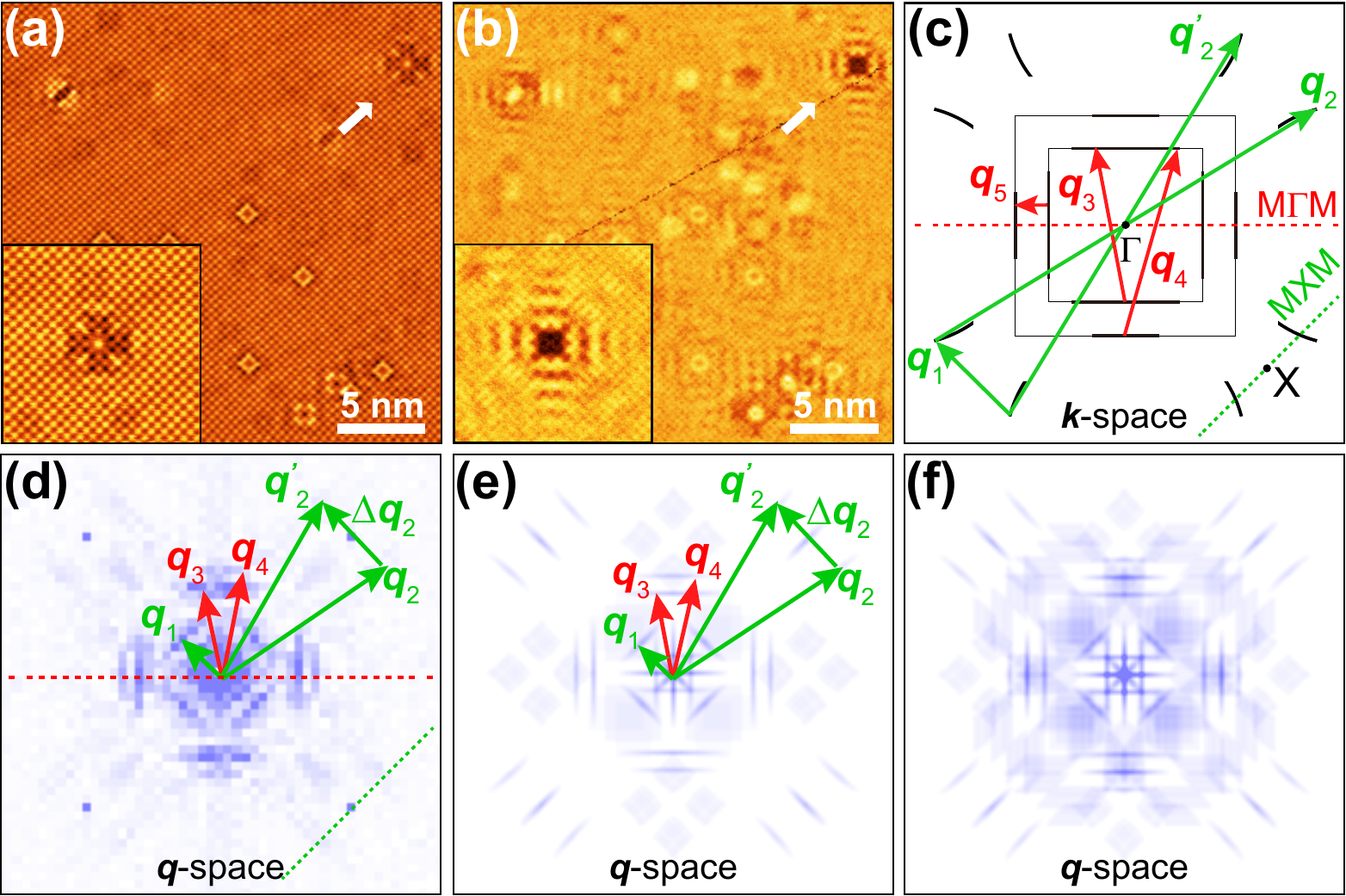}
\caption{(a) A 25 nm $\times$ 25 nm topographic image with bias voltage $V_{\rm b} = 500$ mV. %, $I$ = 1000 pA).
(b) The $dI/dV$ conductance map at $V=500$ mV simultaneously taken with (a).
The insets of (a) and (b) show the  enlarged images around a cross-shaped impurity labelled by a white arrow.
(c) A CCE model in the momentum $\bm k$-space.
(d) The experimental QPI map in the momentum $\bm q$-space using Fourier transform of the local conductance map in the inset of (b).
(e) and (f) are the calculated QPI maps using the CCE model  with and without a selection rule (see text).
The typical scattering wavevectors responsible for the major QPI patterns are shown in (c)-(e).}
\label{fig_n03}
\end{figure}

%%%%%%%%%%%%%%%%%%%%%%%%%%%%%%%%%%%%%%%%%%%%%%%%%%%%%%%%%%%%%%%%%%%%%%%%%%%%%%%%%%%%%%%%% QPI in ZrSiSe

The Fourier transformed scanning tunneling spectroscopy (FT-STS) is next employed to detect
the electronic topology of ZrSiSe, which is resulted directly from its structural symmetry.
With a bias voltage of 500 mV, the topography of a new FOV is displayed in Fig.~\ref{fig_n03}(a),
in which a specific cross-shaped impurity is found on the top right corner.
The $dI/dV$ conductance map simultaneously taken under the same bias voltage
is drawn in Fig.~\ref{fig_n03}(b).
This specific cross-shaped impurity induces a strong elastic scattering, which mixes the electronic
eigenstates of different wavevectors (${\bm k}_i$ and $\bm k_f$) but the same energy. The QPI
is signalled by a standing wave in the LDOS around
the impurity, as shown by an enlarged image in the inset of Fig.~\ref{fig_n03}(b).
The Fourier transform of this local conductance map is drawn in Fig.~\ref{fig_n03}(d).
The QPI patterns in the momentum ${\bm q}$-space can be used to identify the wavevector difference
before and after the elastic scattering (${\bm q}={\bm k}_{f}-{\bm k}_i$),
which helps building the contour of constant energy (CCE).
As shown in Fig.~\ref{fig_n03}(d), the centrally symmetric QPI patterns
can be mainly partitioned into three groups: a  diamond, two concentric  squares, and four sets
of triplet lines. For these pattern groups, we assign their typical scattering wavevectors,
labelled from ${\bm q}_1$ to ${\bm q}_4$ in different colors.
Other QPI patterns cannot be ruled out due to the resolution of our FT-STS map.
In addition, the structure of QPI patterns changes with the scattering impurity
and more discussions are provided in Supplementary Materials.

To interpret the three groups of QPI patterns,
we propose a model CCE
with two groups of $E(\bm k)$ patterns in Fig.~\ref{fig_n03}(c). The first group
includes four pairs of
short arcs around four X points,
contributing to the diamond ($\bm q_1$) and triplet ($\bm q_2$ and $\bm q^\pr_2$)
QPI patterns.
The diamond pattern
results from scattering between the arcs of the same pair,
while the triplet pattern results from scattering between the arcs at the diagonal corners.
The second group consists of two concentric squares of $E(\bm k)$, contributing
to the concentric squares of the QPI patterns ($\bm q_3$ and $\bm q_4$).
The two groups of CCE patterns are similar to those observed in ARPES~\cite{hosen2017tunability}.
However, a key difference is that only the occupied states below the Fermi level are detected
in ARPES. To reproduce the experimental QPI patterns, we introduce
a selection rule that the elastic scattering only occurs between
the  CCE patterns of the same group. The physical mechanism behind this selection
rule is that the two CCE groups belong to the surface and bulk
bands separately~\cite{schoop2016dirac}.
The QPI map calculated based on
the CCE model and the selection rule [Fig.~\ref{fig_n03}(e)]
shares the same major features as those from the experimental measurement
[Fig.~\ref{fig_n03}(d)]. As a comparison, the calculation
without the selection rule [Fig.~\ref{fig_n03}(f)] clearly deviates from the experimental result.

\begin{figure}[tp]
\includegraphics[width=0.5\columnwidth]{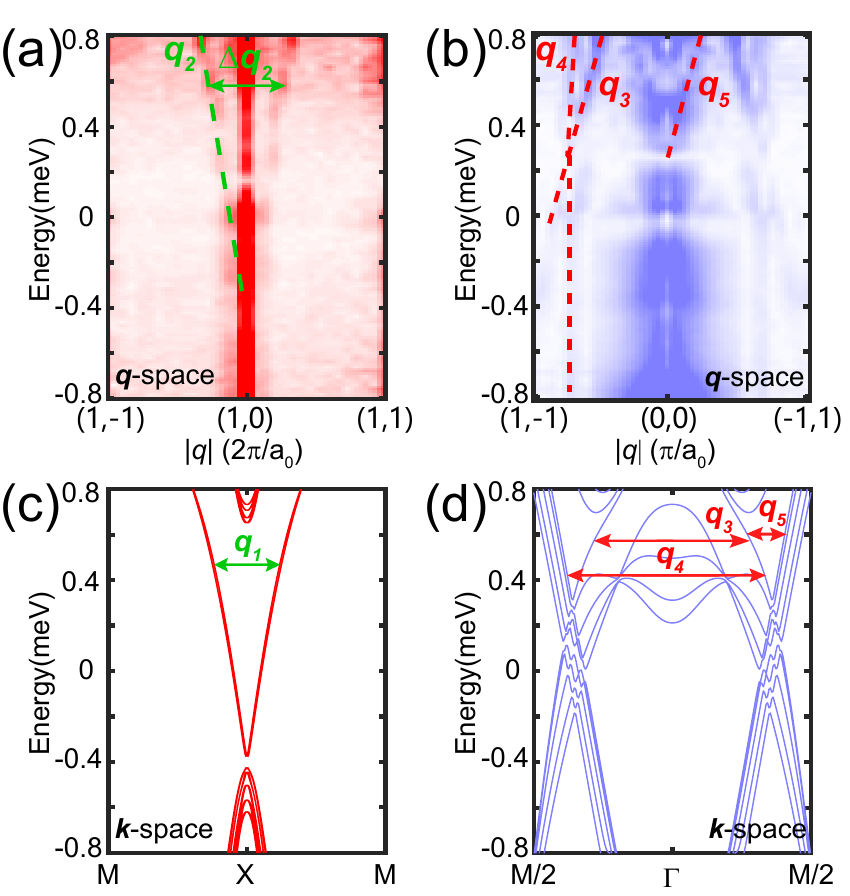}
\caption{(a) and (b) present the QPI energy dispersions along the green and red dashed lines in the $\bm q$-space [Fig.~\ref{fig_n03}(d)].
The energy dispersions of the wavevectors relevant to the electronic topology are highlighted in dashed lines in (a) and (b).
(c) and (d) present the DFT calculation of the slab band structure along the M-X-M and M-$\Gamma$-M directions in the $\bm k$-space.}
\label{fig_n04}
\end{figure}

Next we measure the energy dependent conductance maps around this impurity and study the
energy dispersion relations. Figure~\ref{fig_n04}(a) presents the result of $\bm q(E)$ along the green line in the $\bm q$-space [Fig.~\ref{fig_n03}(d)].
The triplet QPI pattern [Fig.~\ref{fig_n03}(d)] is gradually compressed as the energy approaches
the Fermi level from above.
A linear energy dispersion of the scattering wavevectors, $\bm q_2$ and $\bm q^\pr_2$, is observed.
Their difference, $\Delta \bm q_2 = \bm q^\pr_2-\bm q_2$, vanishes roughly at the bias voltage of $\sim-400$ mV,
indicating a Dirac cone in the electronic band structure. Due to the scattering within the same CCE
group, a simple relation, $\Delta \bm q_2 = 2 \bm q_1$, holds. The energy dispersion of $\bm q_1$ (not shown) follows
the same behavior of $\Delta \bm q_2$, further confirming the existence of the Dirac cone.
In Fig.~\ref{fig_n04}(c), we provide the density functional theory (DFT) calculation of the slab band structure along the M-X-M direction
in the $\bm k$-space.
The predicted Dirac point, if ignoring the small gap
due to the spin-orbital coupling (SOC), is consistent with our experimental measurement.
In addition, this Dirac cone is a surface derived state since it is not observed
in the DFT calculation of the bulk ZrSiSe.

Figure~\ref{fig_n04}(b) presents the result of $\bm q(E)$ along the red line in the $\bm q$-space [Fig.~\ref{fig_n03}(d)].
As the energy approaches the Fermi level from above, the sizes of the two concentric QPI squares
[Fig.~\ref{fig_n03}(d)] are both enlarged,
but with different speeds. These two squares are merged into a single square, indicating the appearance
of a nodal line. The linear energy dispersion is also found for the scattering wavevectors,
$\bm q_3$ and $\bm q_4$. As the amplitude of $\bm q_3$ is increased fast than that of $\bm q_4$,
the crossing point of these two wavevectors leads to an estimation of the nodal line at the
energy of $\sim250$ meV above the Fermi level. In our  CCE model, $\bm q_3$
arises from the scattering within the inner square, while $\bm q_4$ arises from the scattering
between the two opposite sides of the inner and outer squares [Fig.~\ref{fig_n03}(c)].
A possibility causing the change of the $\bm q_3$ and $\bm q_4$ is that the inner square is expanded
and the outer square is shrunk with the decrease of energy. In Fig.~\ref{fig_n04}(d), we present
the DFT calculation of the slab band structure along the M-$\Ga$-M direction in the $\bm k$-space.
The above conjecture of the dispersion relation of the two bands is confirmed theoretically.
In addition to $\bm q_3$ and $\bm q_4$, another scattering wavevector $\bm q_5$ is also observed
due to the scattering between the same sides of the inner and outer CCE squares [Fig.~\ref{fig_n03}(c)].
The dispersion relation of $\bm q_5$ [Fig.~\ref{fig_n04}(b)] confirms the
nodal line at the same energy level as that estimated from the mergence of $\bm q_3$ and $\bm q_4$.
Compared to the surface derived Dirac point at the X point, this nodal line is attributed to
the bulk band structure, as shown by the DFT calculation.
An interesting phenomenon is that the QPI pattern due to the scattering within the outer
CCE square is missing, which is possibly due to the impurity sensitivity on the band scattering.

In summary, we perform a sophisticated STM experiment on a novel 3D topological semimetal, ZrSiSe.
The bias selective topographies allow us to identify the lattice structure of the ZrSe bilayer
with a sub-atomic resolution, which confirms the glide mirror symmetry in ZrSiSe. The QPI technique
in the FT-STS measurement is further applied to extract the electronic structure of ZrSiSe.
By analyzing QPI patterns with assistance of the CCE model and DFT calculation, we determine
the nodal line in the bulk band, at $\sim$ 250 meV above the Fermi level. In addition,
a Dirac point is also determined at $\sim$ 400 meV below the Fermi level.
Compared to an indirect determination in previous ARPES studies, our STM measurement
directly visualizes the topological nodal-line state in ZrSiSe. This method can be generalized
to other nodal-line semimetals in the same family, including two dimensional films.

\begin{acknowledgments}

This work was supported by the National Basic Research Program of China (2014CB921203 and 2015CB921004),
the National Natural Science Foundation of China (NSFC-11374260),
and the Fundamental Research Funds for the Central Universities in China.
F.C., X.L. and Y.S. thank the support of the National Key Research and Development Program (2016YFA0300404)
and the National Nature Science Foundation of China (NSFC-11674326)
and the Joint Funds of the National Natural Science Foundation of China and the Chinese Academy of
Sciences' Large-Scale Scientific Facility (U1432139).

K.B. and Y.F. contribute equally to this work.

\end{acknowledgments}


\begin{thebibliography}{27}
\bibitem{chiu2016classification}
C.-K. Chiu, J. C. Y. Teo, A. P. Schnyder, and S. Ryu,
%\newblock {Classification of topological quantum matter with symmetries}.
\newblock {Rev. Mod. Phys.}
  \textbf{88}, 035005 (2016).


\bibitem{wang2012dirac}
Z. J. Wang, Y. Sun, X.-Q. Chen, C. Franchini, G. Xu, H. M. Weng, X. Dai, and Z. Fang,
%\newblock {Dirac semimetal and topological phase transitions in A$_3$Bi (A = Na, K, Rb)}.
\newblock {Phys. Rev. B}
  \textbf{85}, 195320 (2012).


\bibitem{liu2014discovery}
Z. K. Liu, B. Zhou, Y. Zhang, Z. J. Wang, H. Weng, D. Prabhakaran, S.-K. Mo, Z.-X. Shen, Z. Fang, X. Dai, Z. Hussain, and Y. L. Chen,
%\newblock {Discovery of a three-dimensional topological Dirac semimetal, Na$_3$Bi}.
\newblock {Science}
  \textbf{343}, 864 (2014).


\bibitem{weng2015weyl}
H. M. Weng, C. Fang, Z. Fang, B. A. Bernevig, and X. Dai,
%\newblock {Weyl semimetal phase in noncentrosymmetric transition-metal monophosphides}.
\newblock {Phys. Rev. X}
  \textbf{5}, 011029 (2015).


\bibitem{lv2015experimental}
B. Q. Lv, H. M. Weng, B. B. Fu, X. P. Wang, H. Miao, J. Ma, P. Richard, X. C. Huang, L. X. Zhao, G. F. Chen, Z. Fang, X. Dai, T. Qian, and H. Ding,
%\newblock {Experimental discovery of Weyl semimetal TaAs}.
\newblock {Phys. Rev. X}
  \textbf{5}, 031013 (2015).


\bibitem{xu2015discovery}
S.-Y. Xu, I. Belopolski, N. Alidoust, M. Neupane, G. Bian, C.L. Zhang, R. Sankar, G. Q. Chang, Z. J. Yuan, C.-C. Lee, S.-M. Huang, H. Zheng, J. Ma, D. S. Sanchez, B. K. Wang, A. Bansil, F. C. Chou, P. P. Shibayev, H. Lin, S. Jia, and M. Z. Hasan,
%\newblock {Discovery of a Weyl fermionsemimetal and topological Fermi arcs}.
\newblock {Science}
  \textbf{349}, 613--617 (2015).


\bibitem{inoue2016quasiparticle}
H. Inoue, A. Gyenis, Z. J. Wang, J. Li, S. W. Oh, S. Jiang, N. Ni, B. A. Bernevig, and A. Yazdani,
%\newblock {Quasiparticle interference of the Fermi arcs and surface-bulk connectivity of a Weyl semimetal}.
\newblock {AAAS}
  \textbf{351}, 1184--1187 (2016).



\bibitem{bian2016topological}
G. Bian, T.-R. Chang, R. Sankar, S.-Y. Xu, H. Zheng, T. Neupert, C.-K. Chiu, S.-M. Huang, G. Chang, I. Belopolski, D. S. Sanchez, M. Neupane, C. Liu, B. Wang, C.-C. Lee, H.-T. Jeng, C. Zhang, Z. Yuan, S. Jia, A. Bansil, F. Chou, H. Lin, and M. Z. Hasan,
%\newblock {Topological nodal-line fermions in spin-orbit metal PbTaSe$_2$}.
\newblock {Nat. Commun.}
  \textbf{7}, 10556 (2016).

\bibitem{xu2015two}
Q. Xu, Z. Song, S. Nie, H. Weng, Z. Fang, and X. Dai,
%\newblock {Two-dimensional oxide topological insulator with iron-pnictide superconductor LiFeAs structure}.
\newblock {Phys. Rev. B}
  \textbf{92}, 205310 (2015).


\bibitem{hu2016evidence}
J. Hu, Z. J. Tang, J. Y. Liu, X. Liu, Y. L. Zhu, D. Graf, K. Myhro, S. Tran, C. N. Lau, J. Wei, and Z. Q. Mao,
%\newblock {Evidence of Topological Nodal-Line Fermions in ZrSiSe and ZrSiTe}.
\newblock {Phys. Rev. Lett.}
  \textbf{117}, 016602 (2016).


\bibitem{ali2016butterfly}
M. N. Ali, L. M. Schoop, C. Garg, J. M. Lippmann, R. Lara, B. Lotsch, and S. S. P. Parkin,
%\newblock {Butterfly magnetoresistance, quasi-2D Dirac Fermi surface and topological phase transition in ZrSiS}.
\newblock {Sci. Adv.}
  \textbf{2}, e1601742 (2016).


\bibitem{wang2016evidence}
X. F. Wang, X. C. Pan, M. Gao, J. H. Yu, J. Jiang, J. R. Zhang, H. K. Zuo, M. H. Zhang, Z. X. Wei, W. Niu, Z. C. Xia, X. G. Wan, Y. L. Chen, F. Q. Song, Y. B. Xu, B. G. Wang, G. H. Wang, and R. Zhang,
%\newblock {Evidence of both surface and bulk Dirac bands and anisotropic nonsaturating magnetoresistance in ZrSiS}.
\newblock {Adv. Electron. Mater.}
  \textbf{2}, (2016).


\bibitem{lv2016extremely}
Y.-Y. Lv, B.-B. Zhang, X. Li, S.-H. Yao, Y. B. Chen, J. Zhou, S.-T. Zhang, M.-H. Lu, and Y.-F. Chen,
%\newblock {Extremely large and significantly anisotropic magnetoresistance in ZrSiS single crystals}.
\newblock {Appl. Phys. Lett.}
  \textbf{108}, 244101 (2016).


\bibitem{singha2017large}
R. Singha, A. K. Pariari, B. Satpati, and P. Mandal,
%\newblock {Large nonsaturating magnetoresistance and signature of nondegenerate Dirac nodes in ZrSiS}.
\newblock {Proc. Natl. Acad. Sci.}
  \textbf{114}, 2468 (2017).


\bibitem{hu2017nearly}
J. Hu, Z. J. Tang, J. Y. Liu, Y. L. Zhu, J. Wei, and Z. Q. Mao,
%\newblock {Nearly massless Dirac fermions and strong Zeeman splitting in the nodal-line semimetal ZrSiS probed by de Haas--van Alphen quantum oscillations}.
\newblock {Phys. Rev. B}
  \textbf{96}, 045127 (2017).


\bibitem{sankar2017crystal}
R. Sankar, G. Peramaiyan, I. P. Muthuselvam, C. J. Butler, K. Dimitri, M. Neupane, G. N. Rao, M.-T. Lin, and F. C. Chou,
%\newblock {Crystal growth of Dirac semimetal ZrSiS with high magnetoresistance and mobility}.
\newblock {Sci. Rep.}
  \textbf{7}, 40603 (2017).




\bibitem{schoop2016dirac}
L. M. Schoop, M. N. Ali, C. Stra{\ss}er, A. Topp, A. Varykhalov, D. Marchenko, V. Duppel, S. S. P. Stuart, B. V. Lotsch, and C. R. Ast,
%\newblock {Dirac cone protected by non-symmorphic symmetry and three-dimensional Dirac line node in ZrSiS}.
\newblock {Nat. Commun.}
  \textbf{7}, (2016).


\bibitem{neupane2016observation}
M. Neupane, I. Belopolski, M. M. Hosen, D. S. Sanchez, R. Sankar, M. Szlawska, S.-Y. Xu, K. Dimitri, N. Dhakal, P. Maldonado, P. M. Oppeneer, D. Kaczorowski, F.-C. Chou, M. Z. Hasan, and T. Durakiewicz,
%\newblock {Observation of topological nodal fermion semimetal phase in ZrSiS}.
\newblock {Phys. Rev. B}
  \textbf{93}, 201104 (2016).


\bibitem{hosen2017tunability}
M. M. Hosen, K. Dimitri, I. Belopolski, P. Maldonado, R. Sankar, N. Dhakal, G. Dhakal, T. Cole, P. M. Oppeneer, D. Kaczorowski, F.-C. Chou, M. Z. Hasan, T. Durakiewicz, and M. Neupane,
%\newblock {Tunability of the topological nodal-line semimetal phase in ZrSiX-type materials (X = S, Se, Te)}.
\newblock {Phys. Rev. B}
  \textbf{95}, 161101 (2017).



\bibitem{hoffman2002imaging}
J. E. Hoffman, K. McElroy, D.-H. Lee, K. M. Lang, H. Eisaki, S. Uchida, and J. C. Davis,
%\newblock {Imaging quasiparticle interference in Bi$_2$Sr$_2$CaCu$_2$O$_{8+\delta}$}.
\newblock {Science}
  \textbf{297}, 1148 (2002).



\bibitem{butler2017quasiparticle}
C. J. Butler, Y.-M. Wu, C.-R. Hsing, Y. Tseng, R. Sankar, C.-M. Wei, F.-C. Chou, and M.-T. Lin,
%\newblock {Quasiparticle interference in ZrSiS: Strongly band-selective scattering depending on impurity lattice site}.
\newblock {Phys. Rev. B}
  \textbf{96}, 195125 (2017).


\bibitem{lodge2017observation}
M. S. Lodge, G. Chang, C. Y. Huang, B. Singh, J. Hellerstedt, M. T. Edmonds, D. Kaczorowski, M. M. Hosen, M. Neupane, H. Lin, M. S. Fuhrer, B. Weber, and M. Ishigami,
%\newblock {Observation of Effective Pseudospin Scattering in ZrSiS}.
\newblock {Nano. Lett.}
  \textbf{17}, 7213 (2017).


\bibitem{zheng2017study}
Y. Zheng, Y. Fei, K. L. Bu, W. H. Zhang, Y. Ding, X. J. Zhou, J. E. Hoffman, and Y. Yin,
%\newblock {The study of electronic nematicity in an overdoped (Bi, Pb)$_2$Sr$_2$CuO$_{6+\delta}$
 %superconductor using scanning tunneling spectroscopy}.
\newblock {Sci. Rep.}
  \textbf{7}, 8059 (2017).


\bibitem{lawler2010intra}
M. J. Lawler, K. Fujita, J. Lee, A. R. Schmidt, Y. Kohsaka, C. K. Kim, H. Eisaki, S. Uchida, J. C. Davis, J. P. Sethna, and E.-A. Kim,
%\newblock {Intra-unit-cell electronic nematicity of the high-Tc copper-oxide pseudogap states}.
\newblock {Nature}
  \textbf{466}, 347 (2010).


\bibitem{fujita2014simultaneous}
K. Fujita, C. K. Kim, I. Lee, J. Lee, M. H. Hamidian, I. A. Firmo, S. Mukhopadhyay, H. Eisaki, S. Uchida, M. J. Lawler, E.-A. Kim, and J. C. Davis,
%\newblock {Simultaneous transitions in cuprate momentum-space topology and electronic symmetry breaking}.
\newblock {Science}
  \textbf{344}, 612 (2014).


\bibitem{zeljkovic2015dirac}
I. Zeljkovic, Y. Okada, M. Serbyn, R. Sankar, W. Raman, D. Walkup, W. Zhou, J. Liu, G. Chang, Y. J. Wang, M. Z. Hasan, F. C. Chou, H. Lin, A. Bansil, L. Fu, and V. Madhavan,
%\newblock {Dirac mass generation from crystal symmetry breaking on the surfaces of topological crystalline insulators}.
\newblock {Nat. Mater.}
  \textbf{14}, 318 (2015).


\end{thebibliography}
\end{document}

\title{Supplemental Material for \\ ``Visualization of electronic topology in ZrSiSe by \\scanning tunneling microscopy''}

\author{Kunliang Bu}
\affiliation{Department of Physics, Zhejiang University, Hangzhou 310027, China}
\author{Ying Fei}
\affiliation{Department of Physics, Zhejiang University, Hangzhou 310027, China}
\author{Wenhao Zhang}
\affiliation{Department of Physics, Zhejiang University, Hangzhou 310027, China}
\author{Yuan Zheng}
\affiliation{Department of Physics, Zhejiang University, Hangzhou 310027, China}
\author{Jianlan Wu}
\affiliation{Department of Physics, Zhejiang University, Hangzhou 310027, China}
\author{Fangchu Chen}
\affiliation{Key Laboratory of Materials Physics, Institute of Solid State Physics, Chinese Academy of Sciences, Hefei 230031, China}
\affiliation{University of Science and Technology of China, Hefei 230026, China}
\author{Xuan Luo}
\affiliation{Key Laboratory of Materials Physics, Institute of Solid State Physics, Chinese Academy of Sciences, Hefei 230031, China}
\author{Yuping Sun}
\affiliation{Key Laboratory of Materials Physics, Institute of Solid State Physics, Chinese Academy of Sciences, Hefei 230031, China}
\affiliation{High Magnetic Field Laboratory, Chinese Academy of Sciences, Hefei 230031, China}
\affiliation{Collaborative Innovation Center of Advanced Microstructures, Nanjing 210093, China}
\author{Qiunan Xu}
\affiliation{Beijing National Laboratory for Condensed Matter Physics and Institute of Physics, Chinese Academy of Sciences, Beijing 100190, China}
\author{Xi Dai}
\affiliation{Beijing National Laboratory for Condensed Matter Physics and Institute of Physics, Chinese Academy of Sciences, Beijing 100190, China}
\affiliation{Department of Physics, Hong Kong University of Science and Technology, Clear Water Bay Road, Kowloon, Hong Kong}
\author{Yi Yin}
%\email{yiyin@zju.edu.cn}
\affiliation{Department of Physics, Zhejiang University, Hangzhou 310027, China}
\affiliation{Collaborative Innovation Center of Advanced Microstructures, Nanjing 210093, China}

\maketitle

\setcounter{figure}{0}
\setcounter{equation}{0}
\makeatletter
\renewcommand{\thefigure}{S\@arabic\c@figure}
\renewcommand{\theequation}{S\@arabic\c@equation}
%\maketitle

\normalsize
\section{impurity dependent QPI patterns}
\begin{figure}[tp]
\includegraphics[width=0.6\columnwidth]{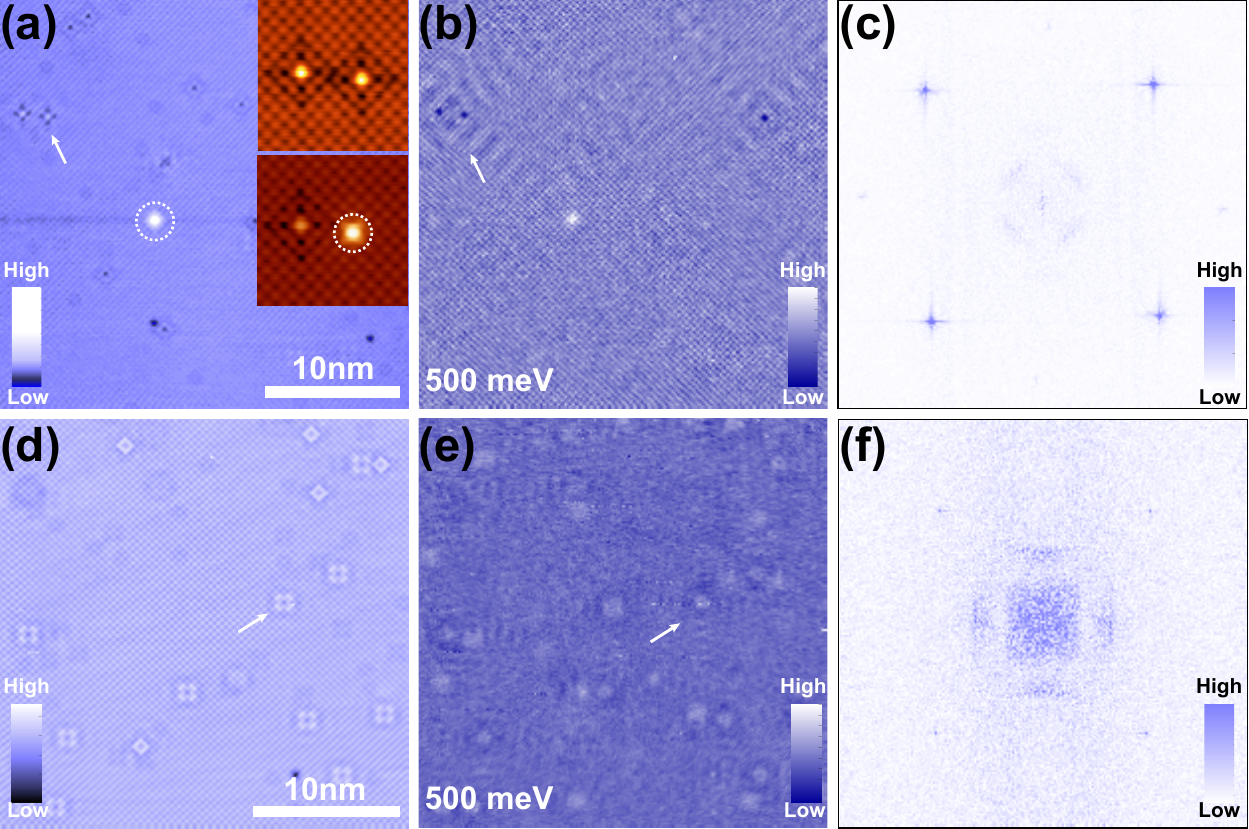}
\caption{(a) 30 nm $\times$ 30 nm topographic image of ZrSiSe ($V$ = 500 mV, $I$ = 400 pA).
Insets show the process in which an adatom (white dashed circle) is generated from a Zr site
defect. Inset images are taken with a junction of $V$ = -500 mV, $I$ = 100 pA.
(b) $dI/dV$ conductance map at 500 meV in the same FOV as (a).
(c) Fourier transform of the conductance map in (b).
(d)-(f) The same as (a)-(c) but with the data taken at another FOV ($V$ = 600 mV, $I$ = 800 pA, 28 nm $\times$ 28 nm).
White arrows denote the defects which dominate the scattering.}
\label{Figure_S1}
\end{figure}

Figure~\ref{Figure_S1}(a) and \ref{Figure_S1}(d) are two chosen topographies, mainly with the Zr
site defects and Se site defects in the field of view (FOV), respectively.
The conductance map simultaneously taken with Fig.~\ref{Figure_S1}(a)
has dominant standing wave patterns around the cross shaped Zr site defects [Fig.~\ref{Figure_S1}(b)].
As shown in Fig.~\ref{Figure_S1}(c), the quasiparticle interference (QPI) in this area
shows a scattering mainly along the direction of the lattice axes.
In contrast, the scattering QPI pattern in Fig.~\ref{Figure_S1}(e)-(f) is along
a direction with an angle of $45^\circ$ to lattice axes. The impurity dependent QPI
dispersion is also observed in ZrSiS (ref. 21), possibly originating from the different
sensitivity of various defects to the band structure.

The cross-shaped Zr site in Fig.~\ref{Figure_S1}(a) is different from that in Fig. 1 in
the main text. In Fig.~\ref{Figure_S1}(a), we also notice an adatom on
top of the surface, as denoted by the white dashed circle. As shown in the
inset of Fig.~\ref{Figure_S1}(a), the lower inset with an adatom is evolved
from the upper inset after several images taken in the same FOV. We thus
conclude that the adatom is generated from the cross shaped Zr site defect.

 \section{Comparison of the QPI dispersion between the experiment and DFT calculation}
\begin{figure}%[tp]
\centering
\includegraphics[width=0.6\columnwidth]{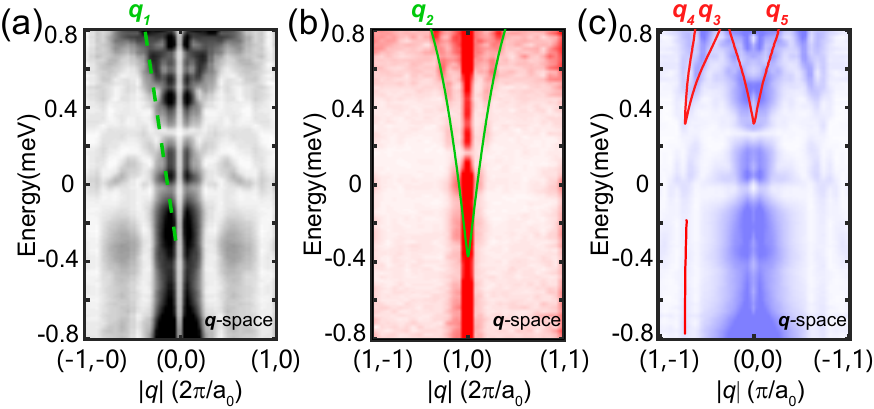}
\caption{(a) QPI dispersion along the X-$\Gamma$-X direction. The $\bm{q_1}$  vector is
highlighted by the green dashed line.
(b) and (c) QPI dispersion along the M-X-M and M-$\Gamma$-M directions. DFT calculations
of the QPI dispersion are superimposed on the experimental data.}
\label{Figure_S2}
\end{figure}

 Figure~\ref{Figure_S2}(a) presents the experimental QPI dispersion along the X-$\Gamma$-X
 direction. The $\bm{q_1}$ vector shows a nearly linear dispersion within the energy range we
 measured. It also forms a Dirac cone at $\sim$~-400 meV, which is consistent with the dispersion
 of $\bm{q_2}$ and the relation between $\Delta\bm{q_2}$ and $\bm{q_1}$. The experimental
 QPI dispersion along the M-X-M and M-$\Gamma$-M directions are reproduced from Fig. 4(a)
 and 4(b), as shown in Fig.~\ref{Figure_S2}(b) and~\ref{Figure_S2}(c). With the density functional theory (DFT) calculation
 of the slab band structure and a selection rule discussed in the main text, we further
 calculate the QPI dispersion along these high-symmetry directions. The calculated QPI
 dispersion is superimposed on its experimental data, as shown
 in Fig.~\ref{Figure_S2}(b) and~\ref{Figure_S2}(c). A quantitative consistence between the
 experiment and the DFT calculation further verify our analysis of the scattering wavevectors
 in the main text.

 \section{Full energy range of the QPI patterns}
 The full evolution of the QPI patterns about the single defect in the inset of Fig. 3(a) is available in Movie S1.